\documentclass{PoS}

\usepackage{times}
\usepackage{mathptmx}
\usepackage{fontenc}

\def\frac#1#2{{#1\over#2}}

\def\half{\ifinner {\scriptstyle {1 \over 2}}%
          \else {\textstyle {1 \over 2}}\fi}

\def\simge{
    \mathrel{\rlap{\raise 0.511ex
        \hbox{$>$}}{\lower 0.511ex \hbox{$\sim$}}}}

\def\simle{
    \mathrel{\rlap{\raise 0.511ex
        \hbox{$<$}}{\lower 0.511ex \hbox{$\sim$}}}}

\def\therefore{
   \setbox0=\hbox{$.\kern.2em.$}\dimen0=\wd0    %
   \mathrel{\rlap{\raise.25ex\hbox to\dimen0{\hfil$\cdotp$\hfil}}%
   \copy0}}

\def\|{\ifmmode\Vert\else \char`\|\fi}

\def\tr{\mathop{\rm tr}\nolimits}       

\def\sc#1{\setbox0=\hbox{$#1$}           
   \dimen0=\wd0                                 
   \setbox1=\hbox{/} \dimen1=\wd1               
   \ifdim\dimen0>\dimen1                        
      \rlap{\hbox to \dimen0{\hfil/\hfil}}      
      #1                                        
   \else                                        
      \rlap{\hbox to \dimen1{\hfil$#1$\hfil}}   
      /                                         
   \fi}                                         %

\def\subrightarrow#1{
  \setbox0=\hbox{
    $\displaystyle\mathop{}
    \limits_{#1}$}
  \dimen0=\wd0
  \advance \dimen0 by .5em
  \mathrel{
    \mathop{\hbox to \dimen0{\rightarrowfill}}
       \limits_{#1}}}                           

\def\vbig#1#2{{\vbigd@men=#2\divide\vbigd@men by 2%
   \hbox{$\left#1\vbox to \vbigd@men{}\right.\n@space$}}}

\usepackage{amssymb}
\usepackage{epsfig}
\usepackage{psfrag}

\PoS{PoS(LAT2005)187}

\title{Large-$N_{f}$ behavior of the Yukawa model: analytic results }

\author{Sergio Caracciolo\\
         Dipartimento di Fisica dell'Universit\`a di Milano \\
         INFN, Sez. di Milano, I-20100 Milano, Italy\\
        E-mail: \email{Sergio.Caracciolo@mi.infn.it}}

\author{\speaker{Bortolo Matteo Mognetti}\\
         Dipartimento di Fisica dell'Universit\`a di Milano \\
        INFN, Sez. di Milano, I-20100 Milano, Italy\\
        E-mail: \email{Bortolo.Mognetti@mi.infn.it}}

\author{Andrea Pelissetto\\
        Dipartimento di Fisica dell'Universit\`a di Roma ``La Sapienza" \\
        INFN, Sezione di Roma I, I-00185 Roma, Italy\\
        E-mail: \email{Andrea.Pelissetto@roma1.infn.it}}

\ShortTitle{Large-$N_{f}$ behavior of the Yukawa model: 
analytic results }

\abstract{We investigate the Yukawa model in which $N_f$ fermions 
are coupled with a scalar field $\phi$ through a Yukawa interaction.
The phase diagram is rather well understood. If the fermions are massless, 
there is a chiral transition at $T_c$: for $T < T_c$ chiral symmetry is 
spontaneously  broken.  At $N_f=\infty$ the transition is mean-field like, 
while, for any finite $N_f$, standard arguments predict Ising behavior. 
This apparent contradiction has been explained by Kogut et al., 
who showed by scaling arguments and Monte Carlo simulations that 
in the large-$N_f$ limit the width of the Ising critical region 
scales as a power of $1/N_f$, so that only mean-field 
behavior is observed for $N_f$ strictly equal to infinity.
We will show how the results of Kogut et al. can be recovered 
analytically in the framework of a generalized $1/N_f$ expansion. 
The method we use is a simple generalization of the method we have 
recently applied to a two-dimensional generalized Heisenberg model.}

\FullConference{XXIIIrd International Symposium on Lattice Field Theory\\
                 25-30 July 2005\\
                 Trinity College, Dublin, Ireland}

\newcommand{\di}{\mathrm{d}}

\newcommand{\ol}{\overline}

\begin{document}

Finite-temperature transitions in quantum field theory models have 
been the object of many theoretical studies. Here we investigate the 
transition in the 2+1 Yukawa model (but the arguments can be generalized to 
generic ($d$+1) models, $d \le 4$) in which a scalar field is coupled to 
$N_f$ degenerate fermions by a Yukawa interaction. As discussed in
Ref.~\cite{KSS-98} this model shows a peculiar behavior for $N_f\to\infty$.
For finite $N_f$, dimensional reduction predicts that the finite-temperature
transition, if continuous, belongs to the two-dimensional 
Ising universality class. On the other hand, for $N_f=\infty$
an explicit calculation gives mean-field behavior \cite{RSY-94}.
These two apparently contradictory results were explained in 
\cite{KSS-98} in terms of a {\it critical-region suppression}. 
A similar behavior was observed in a generalized 
$O(N)$ $\sigma$ model in \cite{CP-02} and an explanation was provided in 
\cite{CMP-05}.
The same techniques 
developed in \cite{CMP-05} can be applied here 
to obtain an analytic description of the 
crossover from mean-field to Ising behavior in the large-$N_f$ limit.

We will investigate the Yukawa model with action \cite{KSS-98}
\begin{eqnarray}
{\cal S}[\phi,\ol \psi,\psi] &=& 
\int d^3 {\mathbf r}\, 
\left[
   {N_{f}\over 2}\big (\partial \phi \big)^2 + N_{f} {\mu \over 2} \phi^2 
   +N_{f} {\lambda  \over 4!} \phi^4 
+\sum_{f=1}^{ N_{f}} 
    \ol \psi_{f} \left(\sc{\partial}+ g\, \phi+M\right) \psi_{f}
\right] ,
\label{modello}
\end{eqnarray}
where the integral is over 
${\mathbb R}^2 \times [0,1/T]$ and we use periodic (antiperiodic) boundary 
conditions for the boson (fermion) in the ``temporal" direction.
We imagine the theory to be somehow regularized with a cutoff that sets
the energy scale. The chosen regularization is not relevant 
for the discussion, and, for instance, the reader may imagine using the 
lattice action with staggered fermions studied in \cite{KSS-98}.

In order to determine the behavior for $N_f = \infty$ we integrate out the 
fermionic degrees of freedom. Starting from (\ref{modello})
we obtain 
\begin{eqnarray}
e^{-N_f \tilde {\cal S}_{\mathrm{eff}}[\phi]} 
&=& \int \prod_{f=1}^{N_{f}}\di\ol\psi_{f}\di\psi_{f}\,
e^{- {\cal S}[\phi,\ol\psi,\psi]}
\label{largeNzbsnc}
\\
\tilde{\cal S}_{\mathrm{eff}}[\phi] &=& 
 {1\over 2}\big(\partial \phi\big)^2 + {\mu\over 2} \, \phi ^2 
+{ \lambda  \over 4!} \phi^4 
- \tr \log \Big(\sc{\partial} + g\, \phi+M \Big)\; .
\label{largeNhbsnc}
\end{eqnarray}
For $N_f = \infty$ we can perform an expansion around 
a translation-invariant saddle point $\ol \phi$. The stationarity condition 
gives the gap equation
\begin{equation}
\overline{\mu} m + {\scriptstyle {1\over6}} \overline{\lambda} m^3 = 
   (m + M) T \sum_{n\in {\mathbb Z}}
   \int^\Lambda {d^2{\mathbf{p}}\over (2\pi)^2} \, 
   {1\over p^2 + (m + M)^2 + (2 n + 1)^2 \pi^2 T^2}.
\label{GE}
\end{equation}
Here we have defined $m \equiv g \overline\phi$, 
$\overline{\mu} \equiv \mu g^{-2}$,
and $\overline{\lambda} \equiv \lambda g^{-4}$. With this choice, the dependence
on $g$ disappears. Note that the gap equation is symmetric under $m \to -m$,
and $M \to - M$ so that it is not restrictive to consider $m\ge 0$.

We first analyze the model for $T=0$. A simple analysis of the gap equation
shows that there are two possibilities for $M=0$. 
There is a critical value ${\ol \mu}_c$ such that 
for $\ol \mu < {\ol\mu}_c$ the relevant solution of the gap
equation is such that $m \not = 0$ (chiral symmetry is broken), 
while in the opposite case we have $m = 0$ (no chiral symmetry breaking). 
For $M\not=0$ the behavior is smooth in the parameters.  In the following we 
will only be interested in the case in which the zero-temperature 
theory shows chiral symmetry breaking. Thus, we shall assume 
$\ol \mu < {\ol \mu}_c$. 
Parameters $\overline{\lambda}$ and $\overline{\mu}$ do not play any additional
role in the model and thus in the following we will not consider explicitly
the dependence on them.

Let us now consider the behavior for $T\not=0$. We restrict the discussion to 
$\overline{\mu} > 0$; in this case symmetry is restored for 
$T\to\infty$ and $M=0$ and
therefore, there exists a value $T_c$ 
such that for $T < T_c$, $m$ is a nonvanishing function of 
$T$ (chiral symmetry is broken), while for $T \ge  T_c$, $m = 0$ 
(chiral symmetry is restored). Explicitly, by using (\ref{GE})
we obtain the relation
\begin{equation}
\overline{\mu} = T_c 
   \sum_{n\in {\mathbb Z}}
   \int^\Lambda {d^2{\mathbf{p}}\over (2\pi)^2} \, 
   {1\over p^2 + (2 n + 1)^2 \pi^2 T^2_c}.
\label{Tc}
\end{equation}
If $M\not=0$, the behavior is smooth 
in $T$. Thus, $T = T_c$, $M = 0$ is a critical point (CP) for model
(\ref{modello}).

Close to the CP one can define a thermal scaling field 
$u_t$ and a magnetic scaling field $u_h$. Using the gap equation
it is easy to see that we can take $u_t = (T - T_c)/T_c$ and 
$u_h = M/T_c$. Then, for $u_t, u_h\to 0$ at fixed $x \equiv u_t u_h^{-2/3}$,
one obtains the mean-field equation of state
\begin{eqnarray}
{m/T_c}
 = {u_h}^{1/3}f(x)\label{MFscale}
\qquad \qquad 
a f(x)^3 + b xf(x)+1 = 0
\end{eqnarray}
with $a>0$. 
From (\ref{MFscale}), if $u_t=0$, we obtain $m \sim {u_h}^{1/3}$, so that
$\delta = \delta_{\rm MF} = 3$; if $u_h =0$, we have 
$m \sim {u_t}^{1/2}$ so that $\beta = \beta_{MF}=1/2$.
Thus, for $N_f = \infty$, the behavior is of mean-field type,
in agreement with previous results \cite{RSY-94}.
Note that the condensate $\Sigma = \langle \ol \psi \psi\rangle$ 
is proportional to $M + m  \approx  {u_h}^{1/3}f(x)$ in the scaling limit.
A completely analogous discussion holds for the staggered lattice model.

In order to perform the $1/N_f$ calculation,
we expand action (\ref{largeNhbsnc}) 
around the saddle-point solution, writing 
$\phi = \overline \phi +\widehat\phi/\sqrt{N_f}$.
We obtain the expansion
\begin{eqnarray}
\tilde{\cal S}_{\rm eff} [\phi_n] &=&
{1\over 2} \sum_n \int {\mathrm{d}^2\mathbf{p} \over (2\pi)^2} \,
\phi_n(\mathbf{p})\,\Delta^{-1}_n(\mathbf{p})\,\phi_{-n}(-\mathbf{p})
\nonumber \\
&&+{1\over 3!\sqrt{N_f}}\sum_{n,m}
\int {\mathrm{d}^2\mathbf{p}\over (2\pi)^2} 
     {\mathrm{d}^2\mathbf{q}\over (2\pi)^2} \,
\tilde V^{(3)}(\mathbf{p},n;\mathbf{q},m;-\mathbf{p}-\mathbf{q},-m-n)
\nonumber\\
&&\qquad\qquad\qquad\qquad\qquad
\phi_n(\mathbf{p})\,\phi_m(\mathbf{q})\,
\phi_{-n-m}(-\mathbf{p}-\mathbf{q})
\nonumber\\
&&+{1\over 4!N_f}\sum_{n,m,t} 
\int {\mathrm{d}^2\mathbf{p}\over (2\pi)^2} 
     {\mathrm{d}^2\mathbf{q}\over (2\pi)^2}
     {\mathrm{d}^2\mathbf{k}\over (2\pi)^2} \,
\tilde V^{(4)}(\mathbf{p},n;\mathbf{q},m;\mathbf{k},t;-\mathbf{p}-\mathbf{q}-\mathbf{k},-m-n-t)
\nonumber\\
&&\qquad\qquad\qquad\qquad\qquad
\phi_n(\mathbf{p})\,\phi_m(\mathbf{q})\,\phi_t(\mathbf{k})\,
\phi_{-n-m-t}(-\mathbf{p}-\mathbf{q}-\mathbf{k}),
\label{theoryphi}
\end{eqnarray}
where the neglected terms are of order $1/N_f^{3/2}$. Here 
$\phi_n({\mathbf p})$ is the Fourier transform of 
$\widehat \phi(x,\mathbf{r})$ ($x\in [0,T^{-1}]$, $\mathbf{r},\mathbf{p}
\in \mathbb{R}^{2}$):
\begin{eqnarray}
\widehat \phi(x,\mathbf{r}) &=& \sum_{n\in \mathbb{Z}} e^{2   \pi i x n T}
\int {\mathrm{d}^2\mathbf{p}\over (2\pi)^2} \,
\phi_n(\mathbf{p}) e^{i\mathbf{p}\cdot \mathbf{r}}.
\label{fourier}
\end{eqnarray}
In the standard approach, $1/N_f$ expansions are obtained by performing 
a perturbative expansion of theory (\ref{theoryphi}) in powers of 
$1/N_f$. This is possible here only far from the CP, since at the CP 
the field $\phi_0$ is massless and the expansion is plagued by
infrared divergences. Indeed, starting from the explicit expression
\begin{eqnarray}
g^{-2} \Delta^{-1}_0(\mathbf{p} = \mathbf{0}) &=&
  \overline{\mu} + {1\over2} \overline{\lambda} m^2 - 
   T \sum_{n\in {\mathbb Z}}
   \int^\Lambda {d^2{\mathbf{q}} \over (2\pi)^2} \, 
   { q^2 + (2 n + 1)^2 \pi^2 T^2 - (m + M)^2 \over 
    [q^2 + (2 n + 1)^2 \pi^2 T^2 + (m + M)^2]^2}
\end{eqnarray}
and using (\ref{Tc}), we find that close to the CP
\begin{eqnarray}
{\Delta}^{-1}_0(\mathbf{0}) &\sim & m^2, (m + M)^2
\label{sing}
\end{eqnarray}
No singularity arises for the other modes $\phi_n$, since 
${\Delta}^{-1}_n(\mathbf{p} = 0) = (2 \pi n T)^2$ at the CP.
The infrared singularity of $\Delta_0(\mathbf{p})$ is of course expected:
${\Delta}^{-1}_n(\mathbf{p})$ is proportional to
 the square of the mass of the boson field that should
vanish at the CP (at the CP the correlation length diverges).

In order to deal with this singularity we use the technique we 
have recently applied in \cite{CMP-05} to a generalized Heisenberg model in two
dimensions. We first integrate out the 
massive modes and obtain an effective action for the zero mode:
\begin{eqnarray}
e^{-{\cal S}_{\rm eff}[\varphi]} &=& \int \prod_{n\neq 0} \di\phi_n 
e^{-\tilde {\cal S}_{\rm eff}[\phi]}.
\end{eqnarray} 
The effective action ${\cal S}_{\rm eff}[\varphi]$ has an expansion 
in powers of $1/N_f$ of the form
\begin{eqnarray}
{\cal S}_{\rm eff}[\varphi] &=& H \varphi(\mathbf{0}) + 
      {1\over 2} \int_\mathbf{p} [K(\mathbf{p}) + r] 
      \varphi(\mathbf{p}) \varphi(-\mathbf{p}) 
\nonumber \\ 
&& +  {\sqrt{u}\over 3!} \int_\mathbf{p} \int_\mathbf{q}
                V^{(3)}(\mathbf{p},\mathbf{q},-\mathbf{p}-\mathbf{q}) 
                \varphi(\mathbf{p}) \varphi(\mathbf{q}) 
                \varphi(-\mathbf{p}-\mathbf{q}) 
\nonumber \\ 
&& + 
      {u\over4!} \int_\mathbf{p} \int_\mathbf{q} \int_\mathbf{s}
         V^{(4)}(\mathbf{p},\mathbf{q},\mathbf{s},
                 -\mathbf{p}-\mathbf{q}-\mathbf{s}) 
           \varphi(\mathbf{p}) \varphi(\mathbf{q}) 
           \varphi(\mathbf{s}) \varphi(-\mathbf{p}-\mathbf{q}-\mathbf{s}),
\label{ccl}
\end{eqnarray}
where $\varphi = a \phi_0 + b$, $u = c/N_f$, and $a$, $b$, $c$ are functions 
of $T$ and $M$ fixed 
by the following normalization conditions:
\begin{equation}
\begin{array}{cll}
& K(\mathbf{p}) = \mathbf{p}^2 + O(\mathbf{p}^4)  &
\qquad\qquad \hbox{\rm for $\mathbf{p} \to 0$}
\\[2mm]
& V^{(4)}(\mathbf{0},\mathbf{0},\mathbf{0},\mathbf{0})=1 &
\\[2mm]
& V^{(3)}(\mathbf{0},\mathbf{0},\mathbf{0})=0 &
\end{array}
\label{norm}  
\end{equation}
for any $T$ and $M$ close to the CP. 
The last condition is not trivial and can be imposed because 
the same property holds at the CP for 
$\tilde{V}^{(3)}(\mathbf{0},0;\mathbf{0},0;\mathbf{0},0)$.
Again, in (\ref{ccl}) we have neglected higher-order terms in 
$1/N_f$.

At this point, the origin of the mean-field--to--Ising crossover is 
quite clear. For $N_f = \infty$ ($u=0$) the zero mode is a 
free field and thus the model shows mean-field behavior. On the other
hand, for finite $N_f$, one must consider the full theory 
(\ref{ccl}), which is nothing but a generalized $\varphi^4$ theory whose 
critical behavior belongs to the Ising universality class. 
Note that in (\ref{ccl}) we have discarded higher-order vertices 
$\varphi^n$ that are multiplied by higher powers of $1/N_f$: 
one can show that they do not play any role in the crossover limit
we shall describe below \cite{CMP-05}.

We wish now to compute the crossover behavior. Formally,
the momentum dependence of the vertices is irrelevant (one may think of the 
formal continuum limit in lattice theories) and thus we can simply consider
(this relation becomes exact in the infrared limit, in which one only
considers the long-distance behavior; for a proof in a lattice framework,
see \cite{CMP-05})
\begin{equation}
{\cal S}_{\rm cont}[\varphi] = 
\int d^2\mathbf{r}\, 
\left[ H \varphi + {\scriptstyle {1\over2}} (\partial\varphi)^2 + 
    {\scriptstyle {1\over2}} r \varphi^2 + 
    {\scriptstyle {1\over 4!}} u \varphi^4\right].
\label{phi4}
\end{equation}
Then, we define $\chi(\mathbf{r}) = \varphi(\mathbf{r}/\sqrt{u})$ and note that 
we can rewrite the action as 
\begin{equation}
{\cal S}_{\rm cont}[\chi] =
\int d^2\mathbf{r}\,
\left[ {H\over u} \chi + {1\over2} (\partial\chi)^2 +
    {r\over 2 u} \chi^2 +
    {1\over 4!}  \chi^4\right].
\end{equation}
In terms of $\chi$, the action is a function of $H/u$ and $r/u$. Then,
consider the zero-momentum connected correlation function $\chi_n$. We have
\begin{eqnarray}
\chi_n &=& 
   \int d^2\mathbf{r}_2\ldots  d^2\mathbf{r}_n\, 
   \langle \varphi(\mathbf{0})\varphi(\mathbf{r}_2) \ldots 
            \varphi(\mathbf{r}_n) \rangle^{\rm conn}
\nonumber \\
   &=&u^{n-1} \int d^2\mathbf{r}_2\ldots  d^2\mathbf{r}_n\, 
   \langle \chi(\mathbf{0})\chi(\mathbf{r}_2) \ldots 
            \chi(\mathbf{r}_n) \rangle^{\rm conn}
      = u^{n-1} f_n(H/u,r/u),
\end{eqnarray}
i.e. $u^{1-n} \chi_n$ is a scaling function of $H/u$ and $r/u$. 
Unfortunately, the derivation is not correct, since we have not taken
into account the presence of the cutoff that breaks scale 
invariance. However, in two dimensions 
(and, in general, for $d < 4$) only a mass renormalization 
(a redefinition of the parameter $r$) is needed in order 
to take care of divergencies. By a proper treatment \cite{BB} one can show that 
there is a function $r_c(u)$ (in two dimensions the determination of this 
function requires only a one-loop computation since the only diverging 
diagram is the tadpole) such that, for $t \equiv r - r_c(u)\to 0$ 
(infrared limit), $H\to0$, and $u\to 0$ (weak-coupling limit), 
the correlation function $\chi_n$ 
satisfies the scaling relation $\chi_n = u^{n-1} f_n(H/u,t/u)$. 
These results extend to theory (\ref{ccl})
although the presence of odd powers of $\varphi$ requires also a 
renormalization of $H$. In \cite{CMP-05} we showed that one can find 
functions $r_c(u)$ and $H_c(u)$, such that, for $t \equiv r - r_c(u)\to 0$,
$h \equiv  H - H_c(u)\to 0$, $u \to 0$ at fixed $h/u$ and $t/u$, 
the correlation function $\chi_n$ scales as 
\begin{equation}
\chi_n\sim u^{1-n} f_n(h/u, t/u).\label{chiccl}
\end{equation}
Functions $f_n(x,y)$ are universal: they do not depend on the explict form
of the vertices and of $K(\mathbf{p})$ and can be computed directly 
in the continuum theory. They are the crossover functions 
that relate mean-field and Ising behavior. Consider, for instance, the 
case $h = 0$. For $t$ fixed and $u \to 0$ we obtain the standard perturbative
expansion; thus, $t/u\to \infty$ corresponds to the mean-field limit.
On the other hand, for $t \to 0$ at $u$ fixed, Ising behavior is 
obtained; $t/u=0$ is the nonclassical limit. By varying $t/u$ 
between 0 and $\infty$ one obtains the full universal crossover behavior. 
The universality of the crossover implies that these functions can be computed 
in completely different settings: one can use field theory 
\cite{BB,PRV}, generalized Heisenberg models such as those considered 
in \cite{CMP-05}, or medium-range models (see \cite{PRV,PV-02} and references 
therein). For instance, crossover curves for the effective exponents 
for $M=0$ can be found in \cite{PRV} (field theory) and in 
\cite{LBB} (in this case the correspondence is $R^2 = b N_f$, where the 
constant $b$ can be computed by matching the corresponding perturbative 
expansions at one loop).

The analysis of \cite{CMP-05} can be simplified in the present model 
since chiral symmetry is preserved by regularization. 
In this case the relations
\begin{equation}
H = 0 \qquad\qquad    V^{(3)}(\mathbf{p},\mathbf{q},\mathbf{r}) = 0
\end{equation}
hold at the CP. They imply $H_c(u) = 0$.

In theory (\ref{ccl}) $r$, $H$, $r_c(u)$, and $u$ are functions 
of $T$, $M$, $N_f$. Thus, the next step consists in determining
how these quantities should scale close to the CP in order 
to keep
$x_t \equiv N_f(r - r_c(N_f))$ and $x_h \equiv  N_f (H-H_c(N_f))$ fixed. A
calculation gives \cite{CMP-05}
\begin{equation}
T_c(N_f) - T_c(N_f=\infty) \approx {a + b \log N_f\over N_f},
\end{equation}
where $a$ and $b$ can be explicitly computed. Moreover, we have
\begin{eqnarray}
{T - T_c(N_f)\over T_c(N_f)} \sim {x_t \over N_f}
\qquad\qquad
{M\over T_c(N_f)} \sim {x_h \over {N_f}^{3/2}}.
\label{scalfiel2}
\end{eqnarray}
Ising behavior is observed for $x_t,x_h \ll 1$. This confirms the 
critical-region suppression predicted in \cite{KSS-98}:
The width of the Ising
critical region scales as $N_f^{-1}$ in the thermal
direction and as $N_f^{-3/2}$ in the magnetic one. For $M = 0$ one can 
also characterize the crossover in terms of the bosonic mass $M_{\rm bos}$.
Indeed, similar arguments (see \cite{PRV}) allow us to predict 
\begin{eqnarray}
{M_{\rm bos}\over T_c} \sim {1\over N_f^{1/2}} 
   f_M[x_t \equiv N_f(T - T_c)/T_c],
\end{eqnarray}
where $f_M(x)$ is a crossover function that behaves as 
$x$ for $x \ll 1$ ($\nu=1$ in the Ising theory) and 
as $\sqrt{x}$ for $x \gg 1$ ($\nu=1/2$ for a Gaussian field).
This relation shows that fixing $x_t$ is essentially equivalent to 
fixing $x_M = N_f^{1/2} M_{\rm bos}/T_c$: Ising behavior is 
observed for $x_M \ll 1$, mean-field behavior in the opposite case.
This is exactly the scaling condition discussed in \cite{KSS-98}.


\providecommand{\href}[2]{#2}\begingroup\raggedright\endgroup

\end{document}